\documentclass[aps,prx,twocolumn,superscriptaddress,a4paper,english,longbibliography]{revtex4-2}
\usepackage{times}
\usepackage{color}
\usepackage[colorlinks=true,urlcolor=blue,citecolor=blue]{hyperref}
\usepackage{url}
\usepackage{soul}
\usepackage{graphicx}
\usepackage{amsmath}
\usepackage{subfigure}
\usepackage{xcolor}
\usepackage{amssymb}
\usepackage{latexsym}
\usepackage{multirow}
\usepackage{hyperref} 
\DeclareGraphicsExtensions{.png,.eps}
\usepackage{float}
\usepackage{ulem}
\usepackage{appendix}
\usepackage{etoolbox}
\usepackage{placeins} 

\usepackage[ruled,linesnumbered]{algorithm2e}

\usepackage{xcolor}
\usepackage[urlcolor=blue]{hyperref}      
\hypersetup{
	colorlinks = true,                    
	citecolor = {blue},
	linkcolor = {purple},
}
\begin{document}	

\title{Universal replication of chaotic characteristics by classical and quantum machine learning}

\author{Sheng-Chen Bai}
\affiliation{Center for Quantum Physics and Intelligent Sciences, Department of Physics, Capital Normal University, Beijing 10048, China}
	
\author{Shi-Ju Ran}
\email[Corresponding author. Email: ]{sjran@cnu.edu.cn}
\affiliation{Center for Quantum Physics and Intelligent Sciences, Department of Physics, Capital Normal University, Beijing 10048, China}

\date{\today}

\begin{abstract}
	Replicating chaotic characteristics of non-linear dynamics by machine learning (ML) has recently drawn wide attentions. In this work, we propose that a ML model, trained to predict the state one-step-ahead from several latest historic states, can accurately replicate the bifurcation diagram and the Lyapunov exponents of discrete dynamic systems. The characteristics for different values of the hyper-parameters are captured universally by a single ML model, while the previous works considered training the ML model independently by fixing the hyper-parameters to be specific values. Our benchmarks on the one- and two-dimensional Logistic maps show that variational quantum circuit can reproduce the long-term characteristics with higher accuracy than the long short-term memory (a well-recognized classical ML model). Our work reveals an essential difference between the ML for the chaotic characteristics and that for standard tasks, from the perspective of the relation between performance and model complexity. Our results suggest that quantum circuit model exhibits potential advantages on mitigating over-fitting, achieving higher accuracy and stability.
\end{abstract}
\maketitle

\textit{Introduction.---} To what extent can a machine learning (ML) model learn from the dynamical data of a chaotic system? Chaos refers to a seemingly disorderly but deterministic behavior of dynamic systems ~\cite{kantz2004nonlinear,strogatz2018nonlinear}, which appears in a wide range of realistic scenarios that are deemed highly non-linear (such as population growth~\cite{may1976simple} and atmosphere~\cite{DeterministicNonperiodicFlow}). A chaotic system possesses several exotic properties, including high sensitivity to the initial conditions and perturbations~\cite{kantz2004nonlinear,strogatz2018nonlinear}. This makes the time series prediction (TPS), particularly after a long-time evolution, almost infeasible.

ML has been demonstrated as a powerful approach for TPS. For instance, neural networks, such as the widely-recognized long short-term memory (LSTM)~\cite{hochreiter1997long,10.1162/neco_a_01199} and the transformer models developed most recently~\cite{NIPS2017_3f5ee243}, exhibit excellent performances in, e.g., weather forecasting~\cite{Bi2023,Chen2023}. The ML-based TSP for chaotic systems involve two key issues: (i) short-term prediction of the system's states~\cite{PhysRevLett.120.024102,10.1063/1.5124926,PhysRevE.102.052203}, and (ii) long-term prediction/replication of the statistical/ergodic dynamical behaviors~\cite{10.1063/1.5010300,10.1063/1.5039508,PhysRevE.98.012215,PhysRevResearch.2.012080,SUN2023113971}. Focusing on the second issue, the previous works mainly attempted to replicate by fixing the hyper-parameters of the dynamical system to be specific values (see, e.g., Refs.~\cite{10.1063/1.5010300,10.1063/1.5039508,PhysRevE.98.012215}). It is of greater challenge and importance to use a single ML model for universally capturing the characteristics with the hyper-parameters varying in a non-trivial range, which remains an open issue.

\begin{figure}[htbp]
	\includegraphics[angle=0,width=1\linewidth]{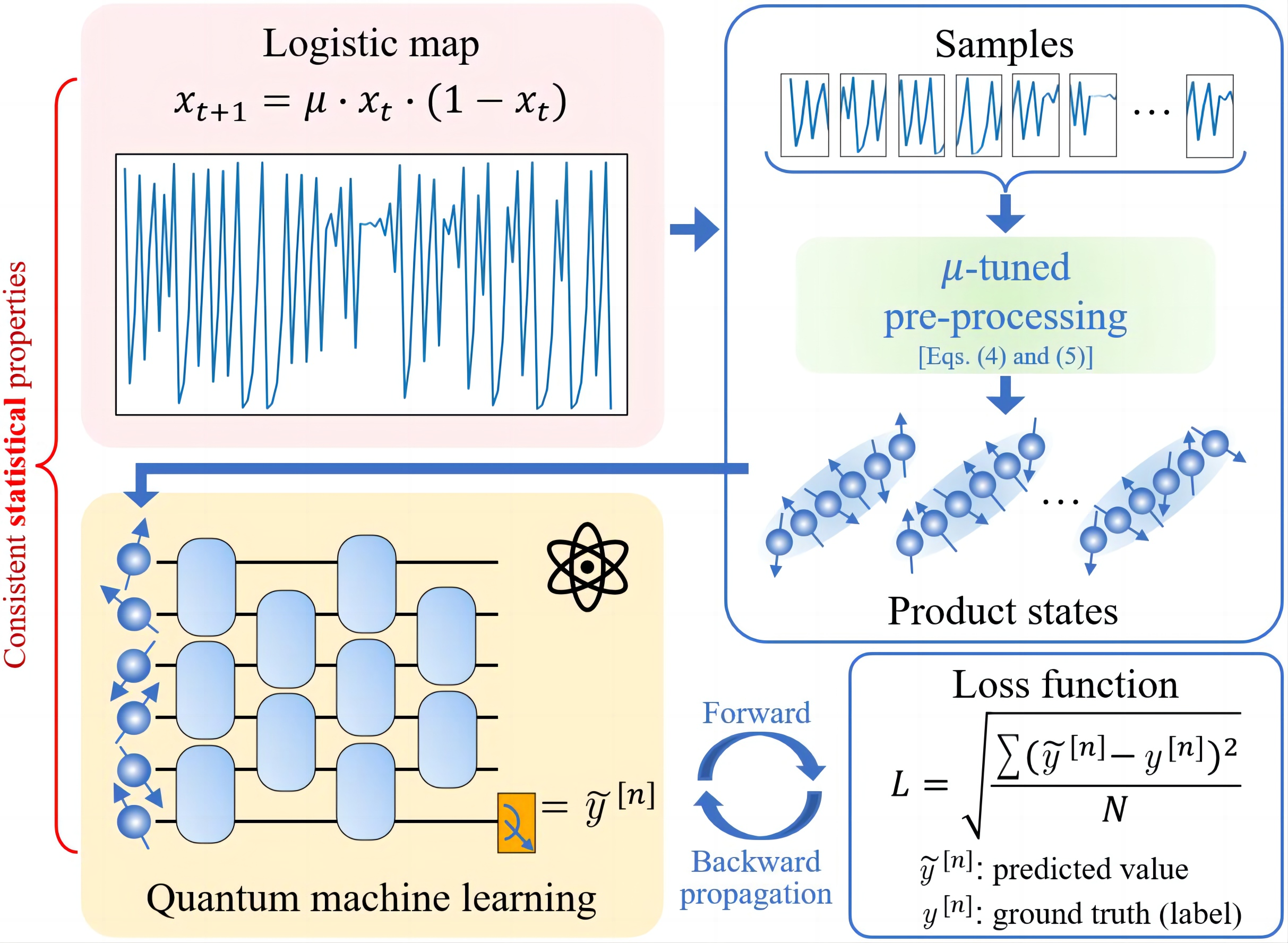}
	\caption{(Color online) The illustration on the main process of time-series prediction by machine learning. The top-left panel shows the one-dimensional Logistic map [Eq.~(\ref{1D logistic})] as an example. By implementing the $\mu$-tuned pre-processing map [see Eqs.~(\ref{eq-preprocess}) and (\ref{eq-featuremap_mu})], a sample (the data of several states) is mapped to a series of vectors as the input of the ML model such as automatically-differentiable quantum circuit (ADQC) illustrated in the bottom-left panel. The ML model is optimized by minimizing the loss function that is taken as the root mean-square error of the one-step-ahead predictions [see the bottom-right panel and Eq.~(\ref{eq-MSE}).}
	\label{fig-idea}
\end{figure}	

In this work, we show that a ML model, which is trained to predict the state one-step-ahead from $M$ latest historic states, can replicate the long-term chaotic characteristics of discrete dynamical systems. We utilize both classical and quantum ML models, namely LSTM~\cite{hochreiter1997long,10.1162/neco_a_01199} and variational quantum circuit~\cite{Peruzzo2014,Cerezo2021} (specifically automatically-differentiable quantum circuit~\cite{zhou2021automatically}, ADQC in short), to learn the one- and two-dimensional Logistic maps~\cite{may1976simple,yuan1983instability,ferretti1988study}. The long-term characteristics, including bifurcation diagram and Lyapunov exponents, are simulated by iterating the ML map for sufficiently many steps. These characteristics, for which the hyper-parameter of the Logistic map varies from stable to chaotic regions, are well replicated by a single ML model.
	
Our results suggest that enhancing the model complexity generally shows no beneficial effects on improving the accuracy of replicating the long-term characteristics due to the high sensitivity to over-fitting. ADQC achieves remarkably higher accuracy and stability than LSTM, particularly for replicating the Lyapunov exponents. These results demonstrate the validity and potential advantage of quantum circuit models on mitigating over-fitting in replicating chaotic characteristics.

\textit{Classical and quantum machine learning for time series prediction.---} A dynamical system can be generally written as a map $x_{t+1}= f(x_{t})$, with $x_{t}$ called the state of system at the (discrete) time $t$. Here, we take the 1D Logistic chaotic map as an example, which is a fundamental nonlinear dynamical system (see the top-left panel of Fig.~\ref{fig-idea}), with wide applications in understanding chaotic phenomena and generating random numbers~\cite{may1976simple}. The map satisfies
\begin{eqnarray}
	f(x_{t}; \mu) = \mu x_{t}(1 - x_{t}),
	\label{1D logistic}
\end{eqnarray}
with $0 \leq \mu \leq 4$ the hyper-parameter.

A ML model is also essentially a map. Considering the prediction of the state one-step-ahead from the latest $M$ historic states, the ML map can be formally written as
\begin{equation}
\tilde{y} = F(x_{1}, x_{2}, ..., x_{M}; \boldsymbol{W}),
\label{eq-F}
\end{equation}
with $\boldsymbol{W}$ denoting all the variational parameters of the ML model. The time series generated by the dynamical system $f$ is used to train the ML model. Accordingly, a ML sample is a piece of the time series with the data of $M$ states, and the label of this sample is the ground truth of the state one-step-ahead given by $f$, i.e., $y = f(x_{M}) = x_{M+1}$. A state in a sample is also called a feature in the language of ML. 

A ML model can be trained by minimizing the prediction error on the so-called training samples. We discretize the hyper-parameter $\mu$ into 50 different values in the range of $2 < \mu \leq 4$. For each value of $\mu$, we generate 2000 samples as the training set, where each sample is obtained by iteratively implementing the dynamical map on a random state for $(M-1)$ times. The variational parameters are optimized by minimizing the root mean-square error (RMSE)
\begin{eqnarray}
	L = \sqrt{\frac{1}{N} \sum_{n} {(\tilde{y}^{[n]} - y^{[n]})}^2,}
	\label{eq-MSE}
\end{eqnarray}
where $\tilde{y}^{[n]}$ denotes the prediction of the ML model for the $n$-th sample, $y^{[n]}$ denotes the ground truth (the label of this sample), and the summation over $n$ goes through the whole training set. The gradient descent method is used to update variational parameters as $\boldsymbol{W} \to \boldsymbol{W} - \eta \frac{\partial L}{\partial \boldsymbol{W}}$ with $\eta$ a small positive constant known as the gradient step or learning rate.

We here consider LSTM and ADQC for ML. LSTM is a recognized classical ML model for TSP. It belongs to the variants of recursive neural networks~\cite{lipton2015critical,salehinejad2018recent} and addresses the issues of vanishing and exploding gradients by introducing the so-called gate functions. ADQC belongs to the variational quantum circuits. Its main advantage is a universal parameterization way of the quantum gates, so that one does not need to specify the types (rotation, phase-shift, controlled-NOT, \textit{etc}.) of gates but just design the structure of the circuit. We use the brick-wall structure as illustrated in bottom-left panel of Fig.~\ref{fig-idea}. The prediction is obtained by measuring the last qubit of the quantum state after implementing the ADQC.

For universally replicating the characteristics of the dynamical system with varying hyper-parameter [say $\mu$ in Eq.~(\ref{1D logistic})] using a single ML model, we propose to introduce a $\mu$-tuned trainable pre-processing on the samples before inputting them to the ML model. Our idea is to map each feature (which is a scalar) to vector by a $\mu$-tuned trainable map, say $x^{[n]}_{t} \to \boldsymbol{v}^{[n, t]} = (v^{[n, t]}_{1},v^{[n, t]}_{2}, \cdots, v^{[n, t]}_{d})$, with $x^{[n]}_{t}$ the $t$-th feature of the $n$-th sample and $d = \dim(\boldsymbol{v}^{[n, t]})$ a preset dimension. In this way, a sample is mapped to a set of vectors. Note that both LSTM and ADQC can take a set of vectors as input. 

The pre-processing map also depends on some variational parameters that will be optimized in the training stage. We here define the pre-processing map as
\begin{eqnarray}
	v^{[n, t]}_{k} = \sum_{i,j} \xi_{i}(x^{[n]}_{t}; \theta) \xi_{j}(\mu^{[n]}; \theta) T_{ijk}.
	\label{eq-preprocess}
\end{eqnarray}
The $(d \times d \times d)$-dimensional tensor $\boldsymbol{T}$ will be optimized when training the ML model. The vectors $\boldsymbol{\xi}(x^{[n]}_{t}; \theta)$ and $\boldsymbol{\xi}(\mu^{[n]}; \theta)$, which are both $d$-dimensional, are obtained by the following map that transforms a scalar to a normalized vector~\cite{NIPS2016_5314b967}. For a given scalar (say $a$), the $j$-th elements of the resulting vector $\boldsymbol{\xi}(a; \theta)$ satisfies
\begin{equation}
	\xi_{j}(a; \theta) = \sqrt{\binom{d-1}{j-1}} \cos \left( \frac{\theta \pi}{2} a \right)^{d-j} \sin \left( \frac{\theta \pi}{2} a \right)^{j-1}.
	\label{eq-featuremap_mu}
\end{equation}
The parameter $\theta$ will also be optimized (independently on $\mu$ and samples) in the training stage. Note since the training samples are obtained from the dynamics taking different values of $\mu$, one should take ${\mu}^{[n]}$ in Eq.~(\ref{eq-preprocess}) to be the corresponding value of $\mu$ for the $n$-th sample. 

The pre-processing map can be generalized to the dynamical system that contains multiple variables and hyper-parameters. We can always use Eq.~(\ref{eq-featuremap_mu}) and map these variables (hyper-parameters) to multiple $d$-dimensional vectors. More details about the dataset and ML models, including pre-processing, forward and backward propagations, prediction, and settings of hyper-parameters can be found in the Appendices.

\begin{figure}
	\includegraphics[angle=0,width=1\linewidth]{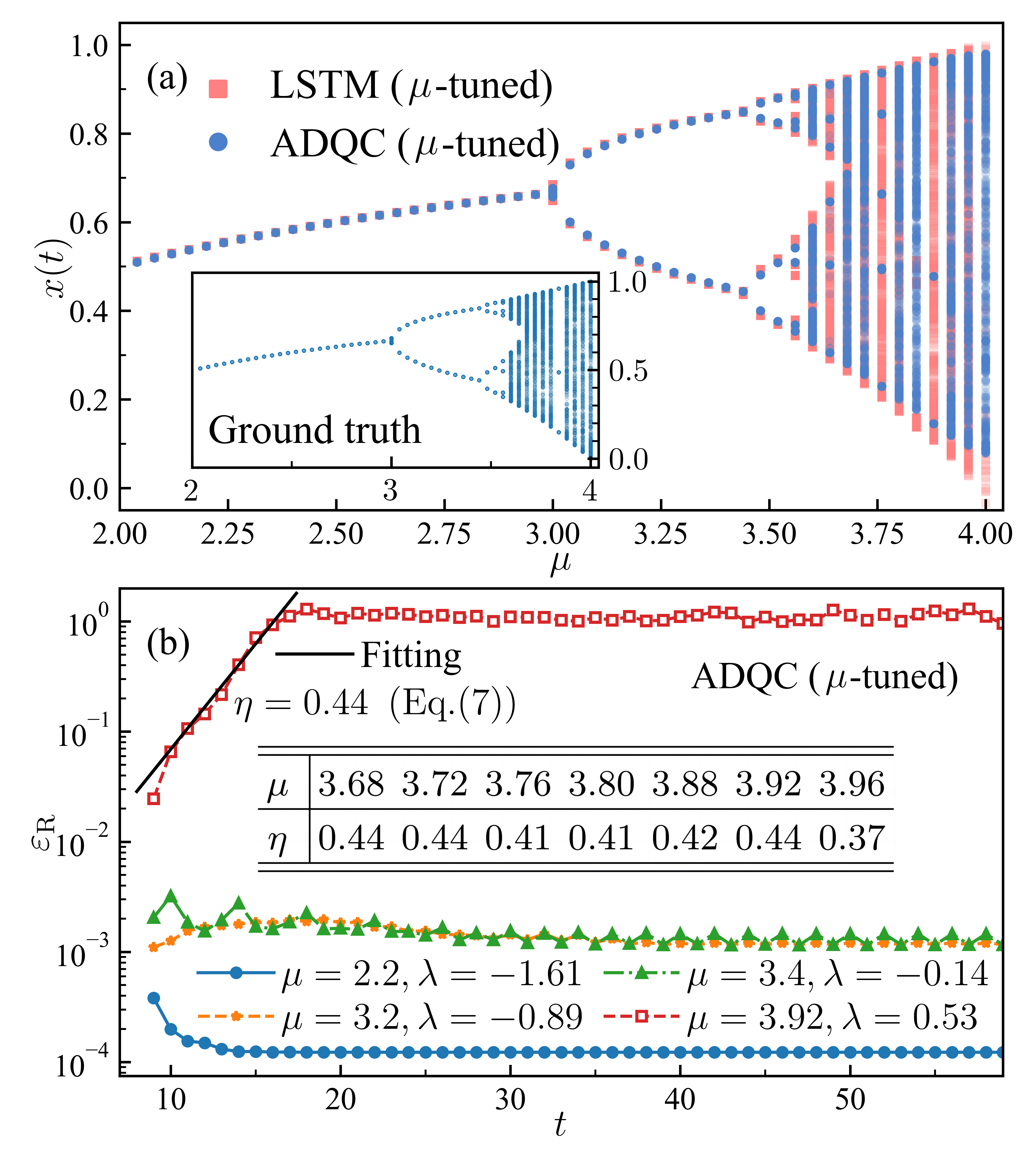}
	\caption{(Color online) (a) The bifurcation diagrams obtained by ADQC, LSTM, and Logistic map (see the inset). The dense of the data points is indicated by the darkness of the colors. (b) The semi-logarithmic plot of the relative $\varepsilon_R$ versus time $t$ in different regions of the bifurcation diagram. We take $\mu= 2.2, 3.2, 3.4$, and $3.92$, which give three negative and one positive values for the Lyapunov exponent. The exponential growth of $\varepsilon_R$ for $\mu=3.92$ is fitted by Eq.~(\ref{eq-LEexp}) with the exponential index $\eta = 0.44$ (see the black solid line). The table in (b) gives the $\eta$ for different values of $\mu$ in the chaotic region.} 
	\label{fig-ability}
\end{figure}	

\textit{Numerical results and discussions.---} Fig.~\ref{fig-ability} demonstrates the ability of LSTM and ADQC on the prediction of the states with different values of $\mu$. In (a), we show the bifurcation diagrams obtained by ADQC and LSTM, which are consistent with the one given by the Logistic map (see the inset). Each data point in the diagrams is obtained by iterating the map (the Logistic map or those of the ML models) for more than 200 times (which is sufficiently large), starting from a random state. We take 500 random states for each value of $\mu$, which are taken differently from those of the training samples. The dense of the data points is indicated by the darkness of the colors. Note we define the samples generated by these initial states as the testing set. Since these samples are not used to train the ML model, the testing set is used reveals the so-called generalization ability, which refers to the power of the ML model on dealing with the unlearned samples.

The similarity between two bifurcation diagrams can be characterized by the peak signal-to-noise ratios (PSNR) $r_{\text{P}}$, which is widely used in the field of computer vision to characterize the similarity between two images~\cite{10.1007/978-3-642-10781-8_37}. We have $r_{\text{P}}=43.40$ and $43.58$ by comparing the ground-truth diagram by Logistic map with the ones by LSTM and ADQC, respectively. Such a consistency between the bifurcation diagrams does not require to accurately predict the states, but requires the validity of accurately replicating the distribution of the states after long-term evolutions. 

In Fig.~\ref{fig-ability}(b), we show the relative error $\varepsilon_{\text{R}} = \frac{1}{N}\sum_{n=1}^{N} \left|(\tilde{y}^{[n]} - y^{[n]}) / y^{[n]} \right|$ by ADQC versus the discrete time $t$ for different values of $\mu$. Note the ML models are always trained to predict the state one-step-ahead. The summation in $\varepsilon_{\text{R}}$ is over the testing test. Different values of $\mu$ are taken so that the system is in the converged ($\mu = 2.2$), bifurcation ($\mu = 3.2$), quadrifurcation ($\mu = 3.4$) and chaotic ($\mu = 3.92$) regions. The values of Lyapunov exponent (LE)~\cite{parker2012practical,moon2008chaotic} are given in the legend, which is calculated as
\begin{eqnarray}
	\lambda= \lim_{T\rightarrow\infty} \frac{1}{T}\sum_{t=1}^{T}\ln\left|\frac{df(x; \mu)}{dx}\right|_{x=x_t}.
	\label{eq-LE}
\end{eqnarray} 
In our numerical simulations, we take $T>200$, which is sufficiently large. 

Our results suggest that for the state prediction in the converged and bifurcation regions (with negative LE), ADQC can well predict the system's state after a long-time evolution, with the relative error $\varepsilon_{\text{R}} \sim O(10^{-3})$. When the system is in a chaotic region (with positive LE), previous work suggests that the state prediction is valid in a short duration characterized usually by the Lyapunov time $T_{\text{LE}} \equiv \lambda^{-1}$~\cite{bezruchko2010extracting}. Our results show that $\varepsilon_{\text{R}}$ grows exponentially with $t$ before it converges to $O(1)$, which obeys
\begin{eqnarray}
	\varepsilon_{\text{R}} \sim e^{\eta t}.
	\label{eq-LEexp}
\end{eqnarray} 
The table in Fig.~\ref{fig-ability}(b) shows the exponential index to be about $\eta \approx 0.4 + O(10^{-1})$. Note the system is chaotic for $3.57 < \mu < 4$~\cite{10.1142/S021797929800051X}. 

\begin{figure}[htbp]
	\centering	\includegraphics[angle=0,width=1\linewidth]{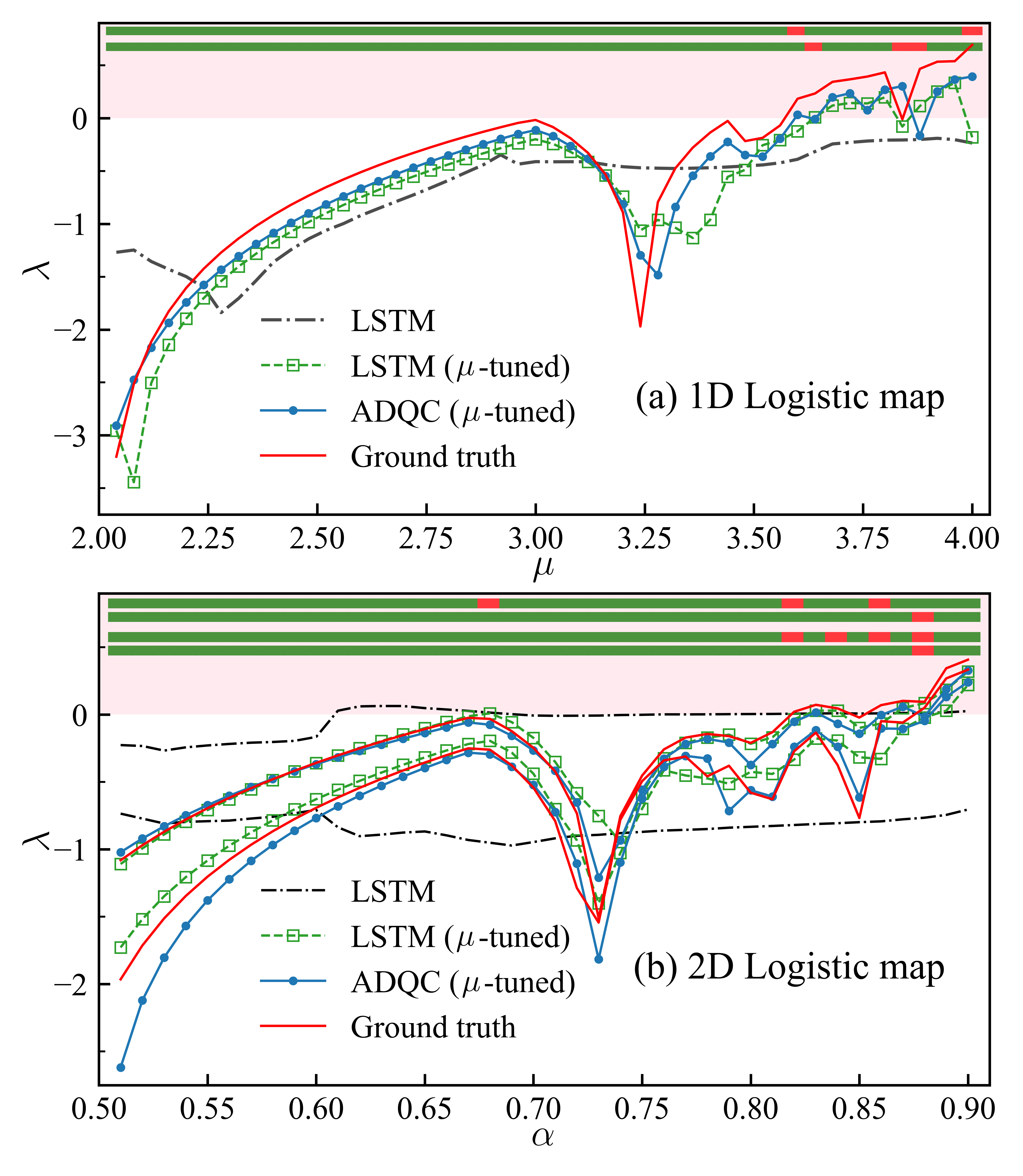}
	\caption{(Color online) The LE's of the (a) 1D and (b) 2D Logistic maps and that of the corresponding ML models (LSTM and ADQC). Each point is the average of five independent simulations. At the top of (a), the two horizontal stripes show whether the LSTM and ADQC correctly (green) or incorrectly (red) give the sign of LE. In (b), the first two stripes at the top show the accuracy on giving the sign of the two LE's by LSTM, and the last two stripes give that by ADQC.}
	\label{fig-LE}
\end{figure}

The above results show the ability of the ML models on replicating the characteristics instead of predicting the states. Such an ability is further demonstrated by comparing the LE of the 1D Logistic map and the ML maps (the trained LSTM and ADQC), as shown in Fig.~\ref{fig-LE} (a). The LE of the ML models is obtained by replacing the differential term in Eq.~(\ref{eq-LE}) by $dF(x_{t-M+1}, ..., x; \boldsymbol{W}] / dx$ at $x=x_t$.

The LE's with varying $\mu$ given by the LSTM and ADQC with the $\mu$-tuned pre-processing (green and blue lines with symbols) are consistent with those given by the Logistic map (the ground truth shown by red solid line). As a comparison, much worse consistence is reached by the standard LSTM without the $\mu$-tuned pre-processing (black dash-dot line). This is a demonstration of the improvement brought by our $\mu$-tuned pre-processing. The two horizontal stripes indicate whether the $\mu$-tuned LSTM and ADQC give the LE with correct (green) or incorrect (red) sign. Generally, the accuracy on replicating the LE's sign is high. The incorrect signs appeal in the chaotic region. 

Similar results are obtained for the two-dimensional (2D) Logistic map [see Fig.~\ref{fig-LE} (b)], which represents a more intricate nonlinear system with richer dynamical behaviors by incorporating a linear coupling term~\cite{yuan1983instability,ferretti1988study}. The two variables $[x_t, x'_t]$ are mapped as 
\begin{eqnarray}
	\begin{array}{c}
		x_{t+1} = 4\mu_1 x_t(1 - x_t) + \beta x'_t    \\
		x'_{t+1} = 4\mu_2 x'_t(1 - x'_t) + \beta x_t 
	\end{array}.
\label{eq-2DLogi}
\end{eqnarray}
We take $\mu_{1} = \mu_{2} = \mu$ and fix $\beta=0.1$ for simplicity. Two LE's are defined. Obviously, replicating the LE's of the 2D Logistic map is much more challenging than the 1D case. Consequently, the LSTM without the $\mu$-tuned pre-processing becomes almost invalid. With the $\mu$-tuned pre-processing, ADQC achieves higher accuracy than LSTM.

\begin{figure}[htbp]	\includegraphics[angle=0,width=1\linewidth]{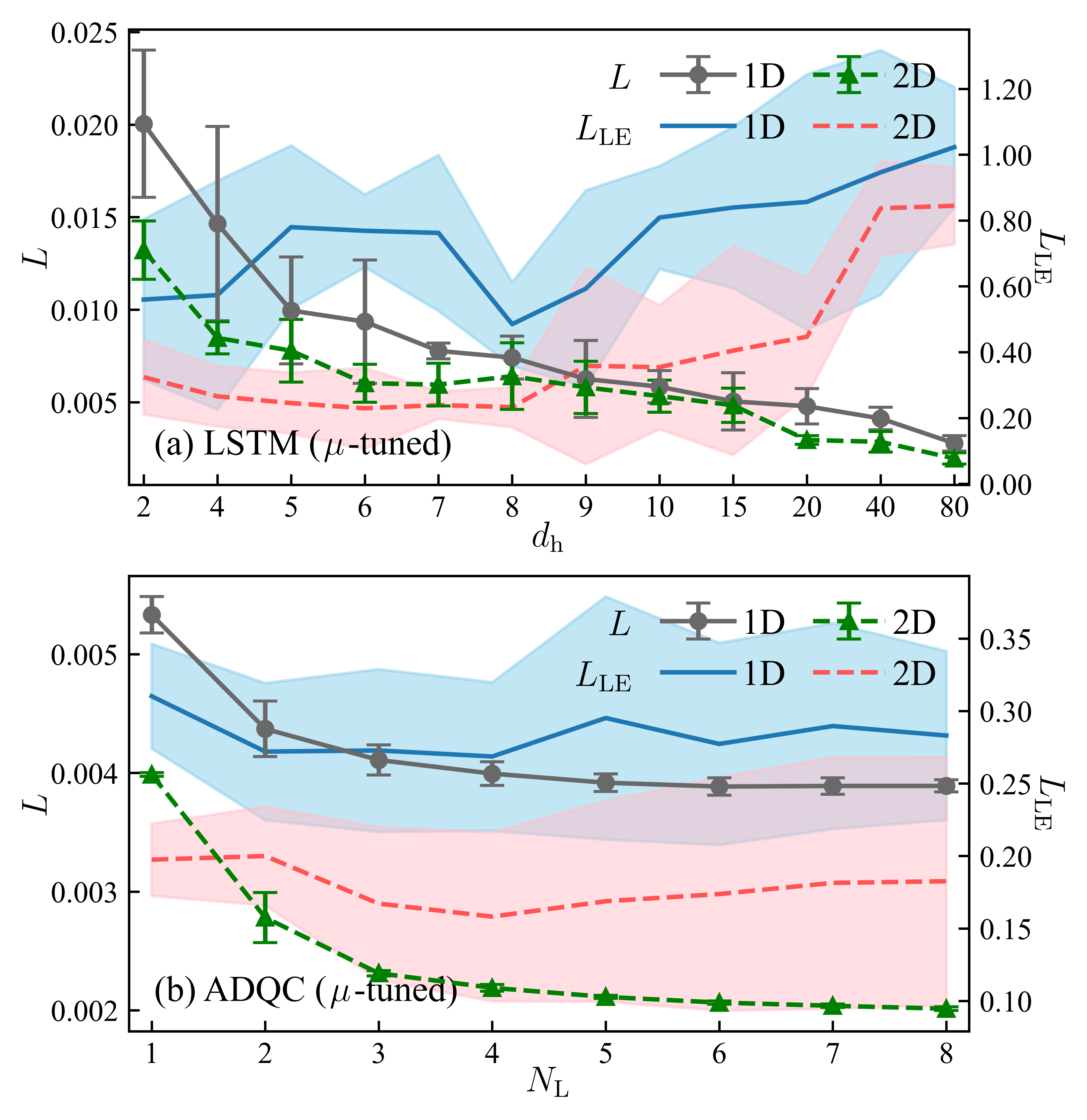}
	\caption{(Color online) The errors of predicting the state one-step-ahead and those of replicating LE for the 1D and 2D Logistic maps. These two errors are characterized by $L$ [Eq.~(\ref{eq-MSE})] and $L_{\text{LE}}$, respectively. In (a), we show the $L$ and $L_{\text{LE}}$ obtained by LSTM versus its hidden dimension $d_{\text{h}}$, and in (b) we show those by ADQC versus its number of layers $N_{\text{L}}$. Each data point is the average on five independent simulations, with the standard deviations demonstrated by the error bars and colored shadows.}
	\label{fig-LE-parameter}
\end{figure}

A usual way to improve the performance of a ML model is to increase its complexity. However, such a way possibly fails in our case. Be noted that though the dynamical system may just contain a handful of hyper-parameters (say one parameter $\mu$ in the 1D Logistic map), leaning its dynamics is a big challenge, and the key factor to determine the performance of a ML model on replicating the long-term characteristics should not be its parameter complexity. We expect high sensitivity to the over-fitting issue.

Fig.~\ref{fig-LE-parameter} supports such an expectation. Increasing the parameter complexity generally lowers the error on the short-term prediction (say predicting the state one-step ahead). For the 1D and 2D Logistic maps, Fig.~\ref{fig-LE-parameter} (a) shows that the RMSE $L$ [Eq.~(\ref{eq-MSE})] obtained by LSTM decreases gradually from about $0.02$ to $0.004$ when increasing the hidden dimension $d_{\text{h}}$ from $2$ to $80$. Similar observation is obtained with ADQC, where $L$ decreases to about $0.002$ (for the 1D Logistic map) or $0.004$ (for the 2D Logistic map) by increasing the number of circuit layers to $N_{\text{L}}=8$. 

The RMSE of LE (denoted as $L_{\text{LE}}$) behaves differently from that of $L$. $L_{\text{LE}}$ is calculated in the same way as $L$ using the definition of RMSE [Eq.~(\ref{eq-MSE})]. For LSTM, the $L_{\text{LE}}$ fluctuates approximately in the range of $0.005< L_{\text{LE}}<0.025$. Increasing $d_{\text{h}}$ of LSTM cannot improve the accuracy on replicating LE. For ADQC, though we fail to see obvious drop of $L_{\text{LE}}$ by increasing $N_{\text{L}}$, $L_{\text{LE}}$ varies much more smoothly approximately in the range of ($0.002<L_{\text{LE}} < 0.005$), which is much lower than that of LSTM. Note all the data in Fig.~\ref{fig-LE-parameter} are computed using the testing set.

Our results suggest that the key of improving the accuracy on the prediction of long-term dynamical characteristics is to developing ML models that better mitigate over-fitting. This is different from the ML tasks such as classification and short-term prediction, where the accuracy can be generally improved by enhancing the model complexity. Our work demonstrates the superior ability of variational quantum circuits on mitigating over-fitting, which is consistent with the previous investigations on different but relevant topics such as image generation~\cite{PhysRevX.8.031012,PhysRevB.99.155131} and model compression~\cite{PhysRevResearch.2.023300,qing2023compressing}. In comparison, LSTM suffers from severe over-fitting issue, though it has been widely recognized as a powerful model with remarkable generalization ability and venial over-fitting issue when dealing with many other ML tasks.

\textit{Summary.---} This work reveals the ability of machine learning (ML) models, which are trained to predict the state one-step-ahead from certain historic data, on replicating the long-term characteristics of discrete dynamical systems. In comparison with the previous works, the characteristics with varying values of hyper-parameters are replicated universally by a single ML model. Our results suggest high sensitivity to the over-fitting issue for the replication of the long-term dynamical characteristics. Taking the one- and two-dimensional Logistic maps as examples, the variational quantum circuit exhibits superior performance on replicating the bifurcation diagram and Lyapunov exponents. Our findings add to the potential advantages of quantum circuit models on achieving high accuracy and stability for the ML tasks that are sensitive to over-fitting.

\textit{Acknowledgment.} SCB is grateful to Qing lv, Peng-Fei Zhou, Yong Qing, Zhang-Xu Chen, Guo-Dong Cheng, Ke Li, Rui Hong, Ying Lu, Yi-Cheng Tang, and Yu-Jia An for helpful discussions. This work was supported in part by NSFC (Grant No. 12004266), Beijing Natural Science Foundation (Grant No. 1232025), Tianjin Natural Science Foundation of China (Grant No. 20JCYBJC00500), and Academy for Multidisciplinary Studies, Capital Normal University.

\appendix
\renewcommand{\thetable}{A\arabic{table}}
\section{Details of dataset and pre-processing}
\label{app-Pre-processing}
As an example, we consider the one-dimensional (1D) Logistic map $f(x_{t}; \mu) = \mu x_{t}(1 - x_{t})$. We take the hyper-parameter $\mu$ to vary from $2$ to $4$, and discretize to $50$ values with an interval of $0.04$. For each value of $\mu$, we randomly generate 3000 training samples and 500 testing samples. Each sample (say $\boldsymbol{x}^{[n]}$) is generated by implementing the dynamical map $f$ on a randomly-taken initial state (say $x^{[n]}_{1}$) for $(M-1)$ times. Therefore, a sample $\boldsymbol{x}^{[n]} = (x^{[n]}_1, \cdots, x^{[n]}_M)$ contains $M$ features with $x^{[n]}_t = f^{t-1}(x^{[n]}_{1})$, where $f^{t-1}$ means to implement $f$ for $(t-1)$ times. 

The ML model is trained to predict the value of $x^{[n]}_{M+1} = f(x^{(n)}_{M}; \mu)$ from this sample. The ground truth of $x^{[n]}_{M+1}$ (generated by the dynamical system itself) is called the label of the $n$-th sample. For different independent simulations, we randomly take $2000$ training samples from the training set for each value of $\mu$ to train the ML models. If not specified, we take $M=8$, and all data such as the bifurcation diagram and the Lyapunov exponents in the main text are computed using the testing set.

A $\mu$-tuned pre-processing is introduced to map each sample to a set of vectors. Specifically, the $t$-th feature of the $n$-th sample $x^{[n]}_{t}$ is mapped to a vector $\boldsymbol{v}^{[n, t]}$ as $x^{[n]}_{t} \to \boldsymbol{v}^{[n, t]} = (v^{[n, t]}_{1},v^{[n, t]}_{2}, \cdots, v^{[n, t]}_{d})$, where the dimension $d = \dim(\boldsymbol{v}^{[n, t]})$ is a preset hyper-parameter. Consequently, a sample is mapped to $M$ vectors.

We define the pre-processing map to depend on $\mu$ and some variational parameters that will be optimized in the training stage. Here, the pre-processing map [also see Eq.~(4) of the main text] is defined as
\begin{eqnarray}
	v^{[n, t]}_{k} = \sum_{i,j} \xi_{i}(x^{[n]}_{t}; \theta) \xi_{j}(\mu^{[n]}; \theta) T_{ijk}.
	\label{app-eq-preprocess}
\end{eqnarray}
The $(d \times d \times d)$-dimensional tensor $\boldsymbol{T}$ and scalar $\theta$ are the variational parameters that are optimized in the training stage.
The vectors $\boldsymbol{\xi}(x^{[n]}_{t}; \theta)$ and $\boldsymbol{\xi}(\mu^{[n]}; \theta)$, which are both $d$-dimensional, are obtained by the following map that transforms a scalar to a normalized vector [also see Eq.~(5) of the main text]. For a given scalar (say $a$), the $j$-th elements of the resulting vector $\boldsymbol{\xi}(a; \theta)$ satisfies
\begin{equation}
	\xi_{j}(a; \theta) = \sqrt{\binom{d-1}{j-1}} \cos \left( \frac{\theta \pi}{2} a \right)^{d-j} \sin \left( \frac{\theta \pi}{2} a \right)^{j-1}.
	\label{app-eq-featuremap_mu}
\end{equation}
Since we train a ML model with different values of $\mu$, one should take ${\mu}^{[n]}$ to be the corresponding value of $\mu$ for the $n$-th sample.

For the 2D logistic map, we take the hyper-parameter $\mu$ to be the discrete values from $0.51$ to $0.9$ with an interval of $0.01$. For each value of $\mu$, we randomly generate 3000 training samples and 500 testing samples. Since it contains two variables ($x_{t}$ and $x'_{t}$), we define the $n$-th sample as $(x^{[n]}_1, x'^{[n]}_1, \cdots, x^{[n]}_M, x'^{[n]}_M)$, and take $M=4$ in our simulations. The pre-processing map is applied similarly to transform the features in a sample to $2M$ vectors.

\section{Automatically-differentiable quantum circuit for time series prediction}	
\label{app-Training ADQC}

The input of an automatically-differentiable quantum circuit (ADQC)~\cite{zhou2021automatically} is usually a $M$-qubit quantum state. With the pre-processing explained above, one maps a sample to a set of vectors, namely from $(x^{[n]}_{1}, \cdots, x^{[n]}_{M})$ to $(\boldsymbol{v}^{[n, 1]}, \cdots, \boldsymbol{v}^{[n, M]})$. These vectors are subsequently mapped to a product state as
\begin{eqnarray}
	\boldsymbol{\psi}^{[n]} = \prod_{\otimes t=1}^M \boldsymbol{v}^{[n, t]}.
	\label{app-eq-IniState}
\end{eqnarray}
In other words, the elements of $\boldsymbol{\psi}^{[n]}$ satisfies $\psi^{[n]}_{s_1, \cdots, s_M} = \prod_{t=1}^M v^{[n, t]}_{s_t}$.

The ADQC represents a unitary transformation (denoted as $\hat{U}$) that maps the input state to the final state that is usually entangled. Formally, we have $\boldsymbol{\Psi}^{[n]} = \hat{U} \boldsymbol{\psi}^{[n]}$ for the state corresponding to the sample $\boldsymbol{x}^{[n]}$, where $\hat{U}$ can be regarded as a $(d^M \times d^M)$-dimensional unitary matrix. For a variational quantum circuit including ADQC, $\hat{U}$ is written as the product of multiple local gates. As illustrated in the bottom-left panel of Fig.~1 of the main text, we here choose the gates to be two-body, which are $(d^2 \times d^2)$-dimensional unitary matrices (denoted as $\{\hat{G}^{[g]}\}$ for $g=1, \cdots, N_{\text{G}}$, with $N_{\text{G}}$ the total number of gates). 

For the 1D Logistic map, the prediction from ADQC is given by the measurement on the final state $\boldsymbol{\Psi}^{[n]}$, satisfying
\begin{eqnarray}
	\tilde{y}^{[n]} = |\Psi^{[n]}_{s_1, \cdots, s_{M-1}, 0}|^2.
\end{eqnarray}
In the quantum language, $\tilde{y}^{[n]}$ is the probability of projecting the last qubit to the spin-up state. For the 1D Logistic map, the predictions of the two variables are defined as
\begin{eqnarray}
	\tilde{y}^{[n]} &=& |\Psi^{[n]}_{s_1, \cdots, s_{M-2}, s_{M-1}, 0}|^2, \\
	\tilde{y}^{[n]} &=& |\Psi^{[n]}_{s_1, \cdots, s_{M-2}, 0, s_{M}}|^2.
\end{eqnarray}
These are the probabilities of projecting the penultimate and last qubits to the spin-up state, respectively.

For the ADQC, each unitary gate is parameterized by a $(d^2 \times d^2)$-dimensional matrix named as latent gates (denoted as $\{\boldsymbol{G}^{[g]}\}$). In other words, the latent gates are the variational parameters of the ADQC. To satisfy the unitary condition of $\{\boldsymbol{G}^{[g]}\}$, the latent gates are mapped to unitary gates by singular value decomposition (SVD) as
\begin{eqnarray}
	&&\boldsymbol{G}^{[g]} \overset{\text{SVD}}\to \boldsymbol{U}^{[g]} \Lambda^{[g]} \boldsymbol{V}^{[g]\dagger}, \\
	&&\boldsymbol{U}^{[g]} \boldsymbol{V}^{[g]\dagger} \to \hat{G}^{[g]}.
\end{eqnarray}

A main advantage of ADQC is that any unitary gate can be parameterized in the same way by a matrix (latent gate) of the same dimensions. Therefore, we only need to specify the structure of the circuit, such as the dimensions of the gates (e.g., the number of spins that one gate acts on and the number of levels for each spin) and how they are connected to each other. For other variational quantum circuit models, one has to additionally specify the types of gates (such as rotational and phase-shift gates). Different types of gates are parameterized in different ways~\cite{10.1002/spe.3039}.

The latent gates (and all other variational parameters) are updated by the gradient decent method in the training stages, as $\boldsymbol{G}^{[g]} \to \boldsymbol{G}^{[g]} - \eta \frac{\partial L}{\partial \boldsymbol{G}^{[g]}}$ with $\eta$ the learning rate. We choose $L$ to be the root mean-square error (see Eq.~(3) of the main text). As the map from latent gates to unitary gates is differentiable, one can use the standard back propagation technique in ML to obtain the gradients.

\section{Lyapunov exponent}
\label{app-LE}

The Lyapunov exponent (LE)~\cite{parker2012practical, moon2008chaotic} is an important measure to describe the chaotic nature of a dynamical system. It reflects the exponential growth rate of small perturbations in the system, quantifying the sensitivity to initial conditions and the amplification of uncertainties. 

Consider a discrete dynamical system with one variable $x_{t+1}= f(x_{t})$ (such as the one-dimensional Logistic map). One can assume an exponential form for the difference between the states with and without a perturbation $\varepsilon$ on the initial state, where one has
\begin{eqnarray}
	\varepsilon e^{T\lambda(x_{0})}=\vert f^{T}(x_{0}+\varepsilon)-f^{T}(x_{0}) \vert,
\end{eqnarray}
where $f^T$ means to recursively implement the map $f$ for $T$ times. LE is defined as the exponential index $\lambda$ in the limit of $\varepsilon \rightarrow 0$ and $T \to \infty$. In practical simulations, LE can be calculated as
\begin{eqnarray}
	\lambda= \frac{1}{T}\sum_{t=1}^{T}\ln\left|\frac{df(x; \mu)}{dx}\right|_{x=x_t},
	\label{eqA-LE}
\end{eqnarray} 
by taking a sufficiently large $T$. In this work, we take $T=264$.

For the 2D case, we employ the QR decomposition method~\cite{geist1990comparison,mcdonald2001error} to compute the LE's. Consider a map with two variables $(x_{t}, x'_{t})$. The map $\boldsymbol{f}$ (which has two components) can be generally written as $x_{t+1} =f_{1}(x_{t}, x'_{t})$ and $x'_{t+1} =f_{2}(x_{t}, x'_{t})$. The Jacobi matrix at the discrete time $t$ is defined as
\begin{eqnarray}
	\renewcommand{\arraystretch}{2}
	\boldsymbol{J}^{(t)} =  \left( \begin{array}{cc} 
		\scalebox{1}{$\dfrac{\partial f_{1}(x, x')}{\partial x}$} &
		\scalebox{1}{$\dfrac{\partial f_{1}(x, x')}{\partial x'}$} \\
		\scalebox{1}{$\dfrac{\partial f_{2}(x, x')}{\partial x}$} &
		\scalebox{1}{$\dfrac{\partial f_{2}(x, x')}{\partial x'}$}
	\end{array} \right)_{x = x_{t}, x'=x'_{t}}.
\end{eqnarray}
The two LE's can be calculates as
\begin{eqnarray}
	\lambda_{k}=  \frac{1}{T}\sum_{t=1}^{T}\ln\left|R^{(t)}_{k, k}\right|,
	\label{eqA-QR}
\end{eqnarray}
with $k=1$ or $2$, and the $(2 \times 2)$ matrix $\boldsymbol{R}^{(t)}$ obtained by implementing QR decomposition $\boldsymbol{J}^{(t)} \to \boldsymbol{Q}^{(t)} \boldsymbol{R}^{(t)}$. We still take $T=264$ in our simulations, which is sufficiently large.

\section{Peak signal-to-noise ratios}

	Peak signal-to-noise ratios (PSNR) $r_{\text{P}}$~\cite{10.1007/978-3-642-10781-8_37} is a measure of the similarity between two images (2D data), which is applied to assess the quality on denoising and compression. Considering two images $\boldsymbol{I}$ and $\boldsymbol{I}'$, the $r_{\text{P}}$ with decibel (dB) as the unit is defined as
\begin{eqnarray}
	r_{\text{P}} = 10 \times \lg\left(\frac{{255}^2}{\frac{1}{WL} \sum_{w=1}^{W} \sum_{l=1}^{L} (I_{wl} - {I'}_{wl})^{2}}\right),
	\label{app-eq-PSNR}
\end{eqnarray}
where $W$ and $L$ are width and length of the images, and 255 is the maximum value of a pixel. In this paper, $r_{\text{P}}$ is computed with the grey-scale images read as 8-bit data. Taking image compression as an example, a PSNR for about $r_{\text{P}} \simeq 40$ dB or larger usually indicates a well-performed compression~\cite{10.1007/978-3-642-10781-8_37}. The larger the $r_{\text{P}}$ is, the closer the two images are to each other.

In this work, each data point in a bifurcation diagram [see Fig.~2(a) of the main text] is obtained by iterating the map (the Logistic map or those of the ML models) for more than 200 times (which is sufficiently large), starting from a random state. We take 500 random states for each value of $\mu$, which are taken differently from those of the training samples. The dense of the data points is indicated by the darkness of the colors. PSNR is used to characterize the similarity between two bifurcation diagrams.

\section{Additional results and hyper-parameters}
	\label{app-details}

Table~\ref{tabelA-lmse} show the RMSE of LE $L_{\text{LE}}$ by LSTM and ADQC on the 1D and 2D Logistic maps. Three observations can be made: 
\begin{enumerate}
\item $L_{\text{LE}}$ can be significantly reduced for both LSTM and ADQC by introducing the $\mu$-tuned pre-processing map;
\item Similar $L_{\text{LE}}$ is achieved with the training and testing sets, indicating sufficient generalization ability on dealing with the unlearned data;
\item With the $\mu$-tuned preprocessing, ADQC achieves much lower $L_{\text{LE}}$ than LSTM, implying that ADQC better mitigating the over-fitting issue than LSTM. This is also consistent with the results reported in the main text (Fig.~4).
\end{enumerate}

\begin{table}[htbp]
\caption{The RMSE of the LE ($L_{\text{LE}}$) on the 1D and 2D Logistic maps obtained by LSTM, ADQC, and those with the $\mu$-tuned pre-processing [dubbed as LSTM ($\mu$) and ADQC ($\mu$)]. Each value of $L_{\text{LE}}$ is obtained by the average of five independent simulations. The standard deviations (\textit{std}) are also provided.}
\begin{tabular*}{8.5cm}{c|c|@{\extracolsep{\fill}}cccc}
	\hline \hline
	\multicolumn{2}{c|}{$L_{\text{LE}}$} & LSTM & LSTM ($\mu$) & ADQC & ADQC ($\mu$) \\ \hline
	\multirow{4}{*}{1D} &Training     & 0.8278 & 0.4845 & 1.1486 & 0.2688 \\ \cline{2-6}
	&\textit{std} & 0.3193 & 0.1280 & 0.0471 & 0.0510 \\ \cline{2-6}
	&Testing      & 0.8273 & 0.4837 & 1.1486 & 0.2686 \\ \cline{2-6}
	&\textit{std} & 0.3190 & 0.1281 & 0.0471 & 0.0512 \\ \hline \hline
	\multirow{4}{*}{2D} &Training     & 0.8050 & 0.2333 & 1.2195 & 0.1588 \\ \cline{2-6}
	&\textit{std} & 0.2320 & 0.0610 & 0.1389 & 0.0584 \\ \cline{2-6}
	&Testing      & 0.8048 & 0.2332 & 1.2189 & 0.1582 \\ \cline{2-6}
	&\textit{std} & 0.2317 & 0.0606 & 0.1388 & 0.0583 \\ \hline \hline
\end{tabular*}
\label{tabelA-lmse}
\end{table}
	
If not specified, the settings of hyper-parameters of LSTM and ADQC used in this work are given in Table~\ref{tabel-para}.

\begin{table}[htbp]
\centering
\caption{The hyper-parameters of ADQC and LSTM for the 1D and 2D Logistic maps. For LSTM, $d_{\text{in}}$ and $d_{\text{out}}$ represent the input and output dimensions, respectively, $L_{\text{SQ}}$ represents the length of sequence, $N_{\text{L}}$ represents the number of layers, and $D_{\text{h}}$ represents the depth. For ADQC, the hyper-parameters are taken in the same way with or without the $\mu$-tuned pre-processing.}
\begin{tabular}{c|l|c}
	\hline \hline
	\multirow{3}{*}{1D} & LSTM      & $d_{\text{in}}=1, \ \ d_{\text{out}}=1,\ \  L_{\text{SQ}}=8, \ \ N_{\text{L}}=1, \ \ D_{\text{h}}=8$ \\ \cline{2-3}
	& LSTM ($\mu$) & $d_{\text{in}}=3, \ \  d_{\text{out}}=1, \ \  L_{\text{SQ}}=8, \ \ N_{\text{L}}=1, \ \ D_{\text{h}}=8$ \\ \cline{2-3}
	& ADQC      & $d=3, \ \  M=8,  \ \  N_{\text{L}}=4$  \\ \hline \hline
	\multirow{3}{*}{2D} & LSTM      & $d_{\text{in}}=2, \ \  d_{\text{out}}=2, \ \  L_{\text{SQ}}=4, \ \ N_{\text{L}}=1, \ \ D_{\text{h}}=8$ \\ \cline{2-3}
	& LSTM ($\mu$) & $d_{\text{in}}=3, \ \  d_{\text{out}}=2, \ \  L_{\text{SQ}}=8, \ \ N_{\text{L}}=1, \ \ D_{\text{h}}=8$ \\ \cline{2-3}
	& ADQC      & $d=3, \ \  M=8, \ \  N_{\text{L}}=4$  \\ \hline \hline
\end{tabular}
\label{tabel-para}
\end{table}


\FloatBarrier 




\normalem
%

\end{document}